\documentclass[10pt]{iopart}
\usepackage{iopams}
\usepackage{subfigure}
\usepackage{tensind,graphicx,gensymb,bbm,wasysym}
\usepackage{mathptmx}      
\usepackage{color}
\usepackage{url}

\newcommand{\diff}{\mathrm{d}}

\begin{document}
\title{Visual exploration of 2D autonomous dynamical systems}

\author{Thomas M{\"u}ller}
\address{
  Visualisierungsinstitut der Universit{\"a}t Stuttgart (VISUS),\\
  Allmandring 19, 70569 Stuttgart, Germany
}
\ead{Thomas.Mueller@visus.uni-stuttgart.de}

\author{Filip Sadlo}
\address{
  Visualisierungsinstitut der Universit{\"a}t Stuttgart (VISUS),\\
  Allmandring 19, 70569 Stuttgart, Germany
}
\ead{Filip.Sadlo@visus.uni-stuttgart.de}

\begin{abstract}
  In an introductory course on dynamical systems or Hamiltonian mechanics, vector field diagrams are a central tool to show a system's qualitative behaviour in a certain domain. 
  Because of their low sampling rates and the involved issues of vector normalization, these plots give only a coarse insight and are unable to convey the vector field behaviour at locations with high variation, in particular in the neighborhood of critical points.
  Similarly, automatic generation of phase portraits based on traditional sampling cannot precisely capture separatrices or limit cylces.
  In this paper, we present \emph{ASysViewer}, an application for the interactive visual exploration of two-dimensional autonomous dynamical systems using line integral convolution techniques for visualization, and grid-based techniques to extract critical points and separatrices.
  \emph{ASysViewer} is addressed to undergraduate students during their first course in dynamical systems or Hamiltonian mechanics.\\[0.5em]
  (Some figures may appear in colour only in the online journal)
\end{abstract}



\section{Introduction}\label{sec:intro}
The qualitative behaviour of two-dimensional autonomous dynamical systems is traditionally explored visually by means of either a vector field diagram or as a phase portrait that shows critical points, several typical orbits, and separatrices. 
Both techniques, however, have the disadvantage that they only coarsely sample the region of interest. 
Furthermore, traditional automatic generation of these plots, in general, cannot precisely capture the separatrices or reproduce limit cylces.
A continuous visualization of the phase space can be achieved using line integral convolution (LIC), which is a well-known technique for vector field visualization introduced by Cabral and Leedom~\cite{cabral1993} in 1993. 
Unfortunately, up to now, this technique has not found its way into lectures about dynamical systems because of the lack of software that can produce high-quality phase images interactively. 

The aim of this article is to present how visualization and numerical techniques can help in the visual and interactive exploration of two-dimensional autonomous dynamical systems which might be helpful in undergraduate courses on classical Hamiltonian mechanics as well as on dynamical systems in general.
For that, we mainly use line integral convolution techniques, integral curves, slicing, and topology extraction, which we bring to the graphics board.

A detailed introduction to general dynamical systems can be found in the standard literature, e.g., Lynch~\cite{lynch} or Wiggins~\cite{wiggins}.
A state of the art report in flow visualization using dense and texture-based techniques, in particular various LIC techniques, was given by Laramee et al.~\cite{laramee2004}.


The structure of this paper is as follows. 
In section~\ref{sec:autonsys}, we briefly review 2D autonomous systems and the classification of critical points.
In section~\ref{sec:vistech}, we discuss the standard vector field diagram and how it can be replaced by line integral convolution visualization. Furthermore, we detail our topology extraction method. In section~\ref{sec:application}, we present our application and give several examples in section~\ref{sec:examples}.

Our software \emph{ASysViewer} is based on the cross-platform framework Qt~\cite{qt} and the Open Graphics Library (OpenGL)~\cite{opengl}. The source code and Windows binaries are available from \url{http://go.visus.uni-stuttgart.de/asysviewer}.


\section{Autonomous dynamical systems}\label{sec:autonsys}
A two-dimensional autonomous dynamical system is defined by two coupled first-order differential equations
\begin{equation}
  \dot{x} = P(x,y),\qquad \dot{y} = Q(x,y),
  \label{eq:system}
\end{equation}
where a dot represents differentiation with respect to time, $t$, and $P$ and $Q$ are two (arbitrary) $C^1$-differentiable functions of $x$ and $y$ in an open subset $\Omega\subset\mathbbm{R}^2$. 
Both, $x$ and $y$ themselves are functions of $t$. 
Because of the theorem of existence and uniqueness, solutions of Eq.~(\ref{eq:system}) do not cross. 
Hence, we can interpret the system of equations (\ref{eq:system}) geometrically as a two-dimensional vector field with a vector $\vec{f}(x,y)=\left(P(x,y),Q(x,y)\right)^T$ attached at each point $(x,y)\in\Omega$. 
The trace of a solution $\vec{\sigma}(t,x_0,y_0)=\left(\sigma_x(t,x_0,y_0),\sigma_y(t,x_0,y_0)\right)^T$ of the coupled system (\ref{eq:system}) with initial values $(x_0,y_0)$ and parameter $t$ is called integral curve (trajectory, orbit, flow line, characteristic line, or streamline), where $\vec{\sigma}$ is always tangential to the vector $\vec{f}$ at the current point $(x,y)$. 
Contour lines for which $\diff{y}/\diff{x}=\dot{y}/\dot{x}=\mbox{const}$ are called \emph{isoclines}.

\begin{figure}[ht]
  \centering
  \includegraphics[width=0.5\textwidth]{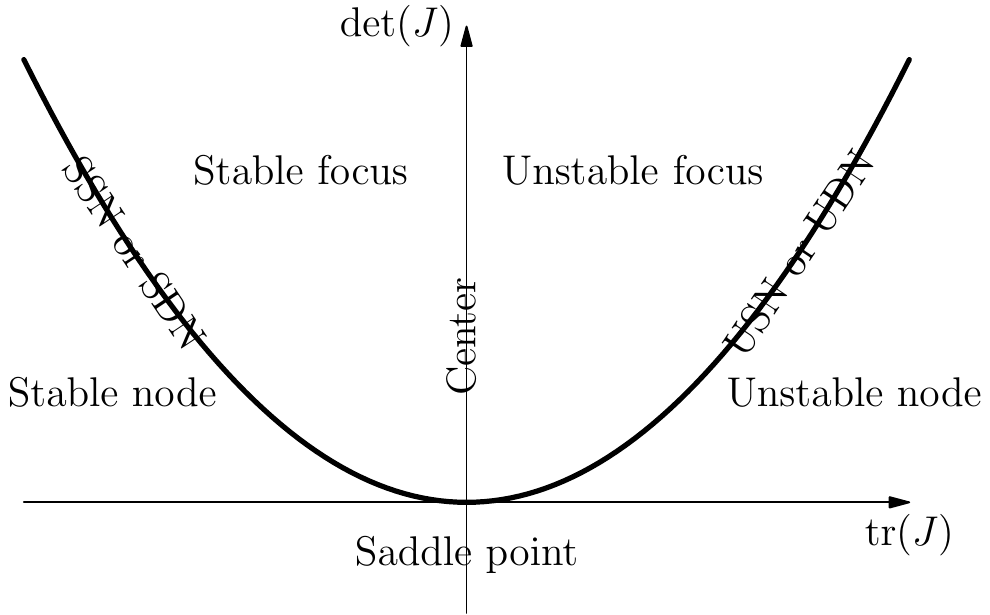}
  \caption{Classification of critical points depending on the trace and the determinant of the Jacobian $J$ (see also Lynch\cite{lynch}). 
  The parabola is given by $\det(J)=(\tr(J))^2/4$, see Eq.~(\ref{eq:lambdas}).}
  \label{fig:classification}
\end{figure}

To obtain a qualitative view of the vector field, a \emph{phase portrait} that consists of critical points and several characteristic lines and separatrices is used.
\emph{Stationary points} (also called fixed or equilibrium points) are defined by $\dot{x}=\dot{y}=0$. 
\emph{Critical points} are isolated stationary points, i.e., stationary points where the determinant of the Jacobian does not vanish, $\det(J)\neq 0$.
The behaviour of the system (\ref{eq:system}) in a close neighborhood of a critical point $(x_c,y_c)$ is characterized by the eigenvalues $\lambda_{1,2}$ of the Jacobian
\begin{equation}
  J = \left(\!\!\begin{array}{cc} J_{11} & J_{12}\\ J_{21} & J_{22}\end{array}\!\!\right) =  \left(\!\!\begin{array}{cc} \partial_xP & \partial_yP \\ \partial_xQ & \partial_yQ\end{array}\!\!\right)\bigg|_{x_c,y_c},
\end{equation}
which follows from linearizing the system (\ref{eq:system}) with a Taylor series expansion,
\begin{equation}
 \dot{X} = X\frac{\partial P}{\partial x}\bigg|_{x_c,y_c} + Y\frac{\partial P}{\partial y}\bigg|_{x_c,y_c},\qquad \dot{Y} = X\frac{\partial Q}{\partial x}\bigg|_{x_c,y_c} + Y\frac{\partial Q}{\partial y}\bigg|_{x_c,y_c},
\end{equation}
and $X=x-x_c$, $Y=y-y_c$. 
Then, the eigenvalues $\lambda_{1,2}$ can be determined by means of the characteristic equation $\det(J-\lambda\mathbbm{1}_2)=0$,
\begin{equation}
 \lambda_{1,2} = \frac{1}{2}\left[\tr(J)\pm\sqrt{\left(\tr(J)\right)^2-4\det(J)}\right]
 \label{eq:lambdas}
\end{equation}
with trace, $\tr(J)=J_{11}+J_{22}$, determinant $\det(J)=J_{11}J_{22}-J_{12}J_{21}$, and $\mathbbm{1}_2$ being the identity matrix in two dimensions. 
If the real part of the eigenvalues is nonzero, the critical point is called hyperbolic (structurally stable, i.e., robust against perturbations), and, according to Hartman's theorem, the phase portrait of the original system in the neighborhood of this critical point resembles that of the linearized system.
Figure~\ref{fig:classification} shows the classification of critical points depending on the values of $\tr(J)$ and $\det(J)$.
If $\det(J)<0$, the critical point can only be a saddle point.
If $\det(J)>0$, it depends on $\tr(J)$ whether the critical point is stable or unstable, i.e., in which way orbits approach or move away from the critical point, respectively.

\begin{figure}[ht]
  \subfigure[unstable node]{\includegraphics[width=0.325\textwidth]{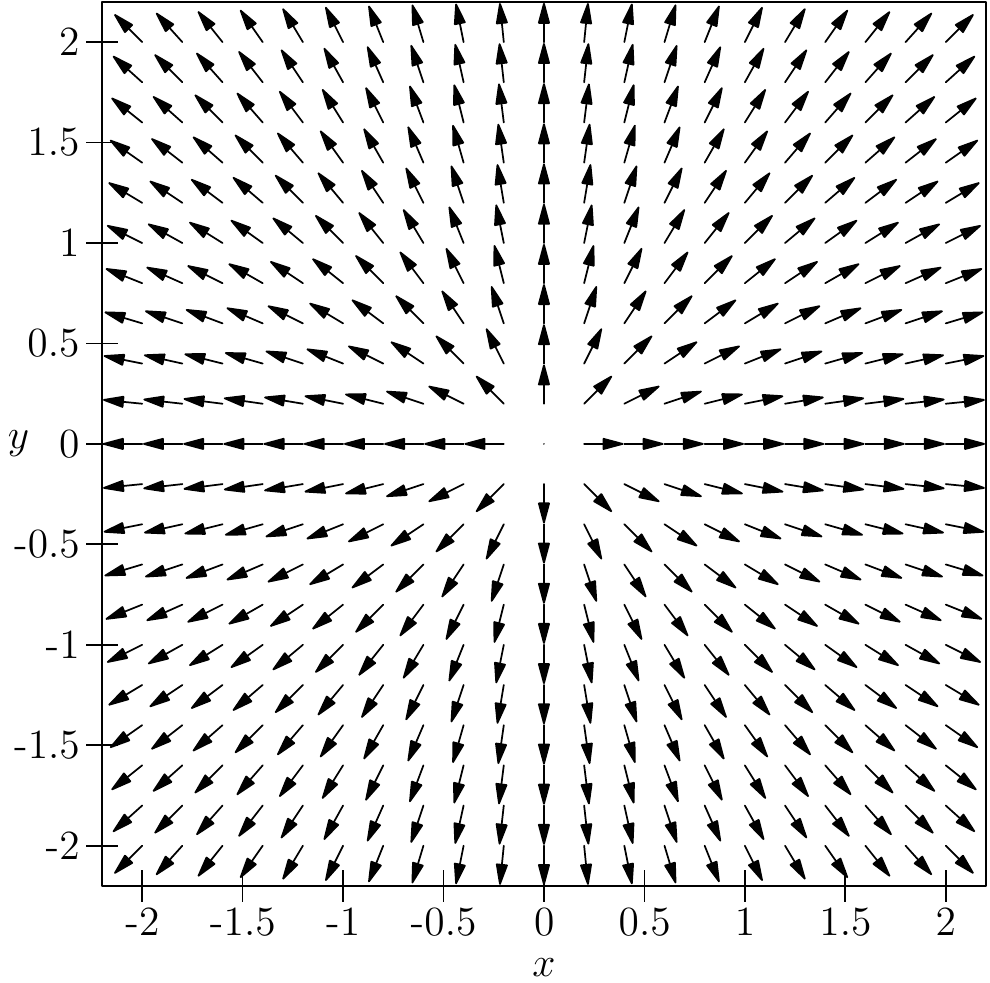}}\hfill
  \subfigure[stable node]{\includegraphics[width=0.325\textwidth]{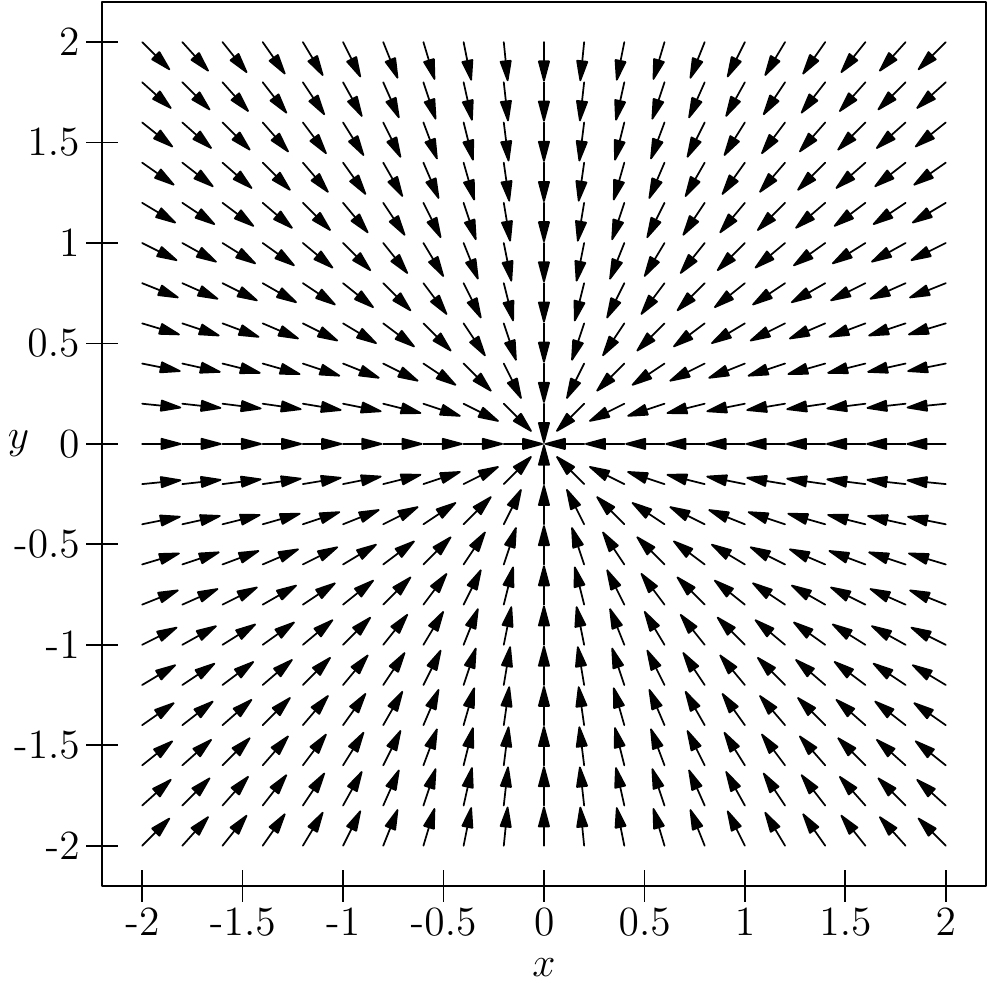}}\hfill
  \subfigure[saddle]{\includegraphics[width=0.325\textwidth]{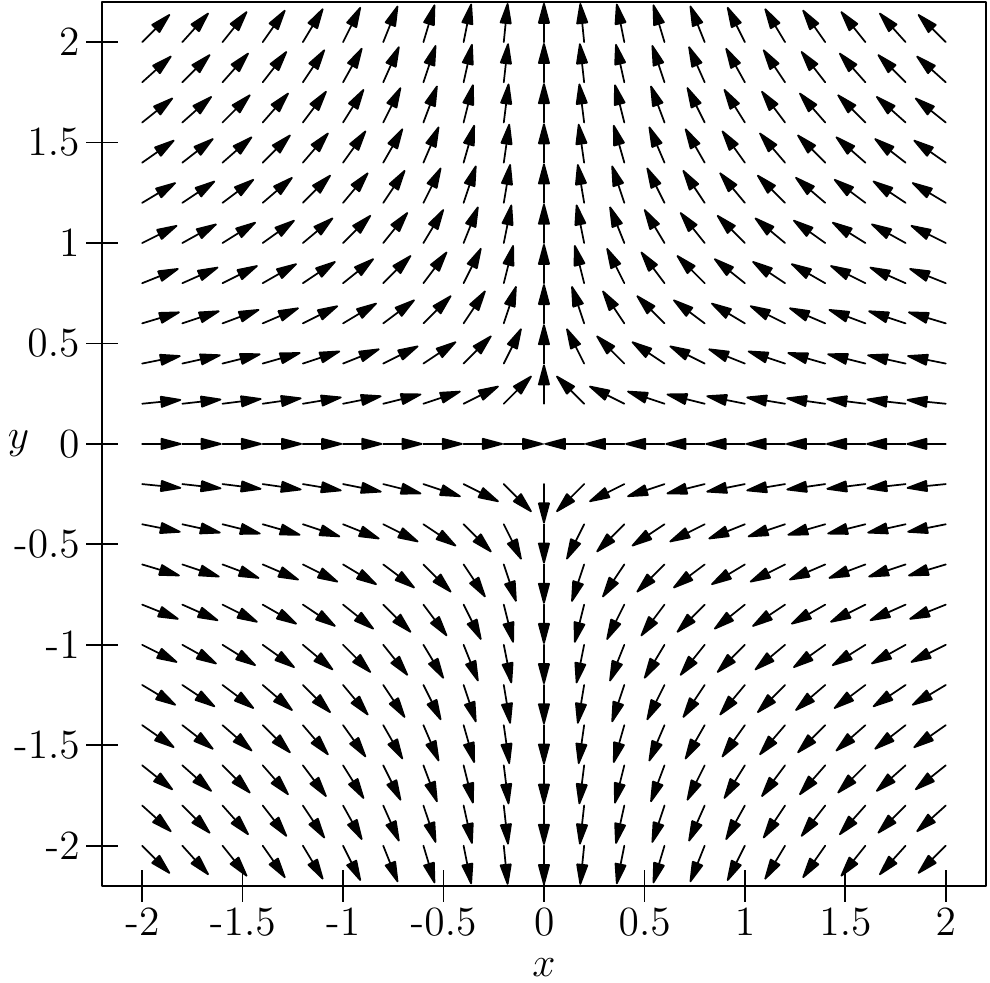}}\\
  \subfigure[unstable focus]{\includegraphics[width=0.325\textwidth]{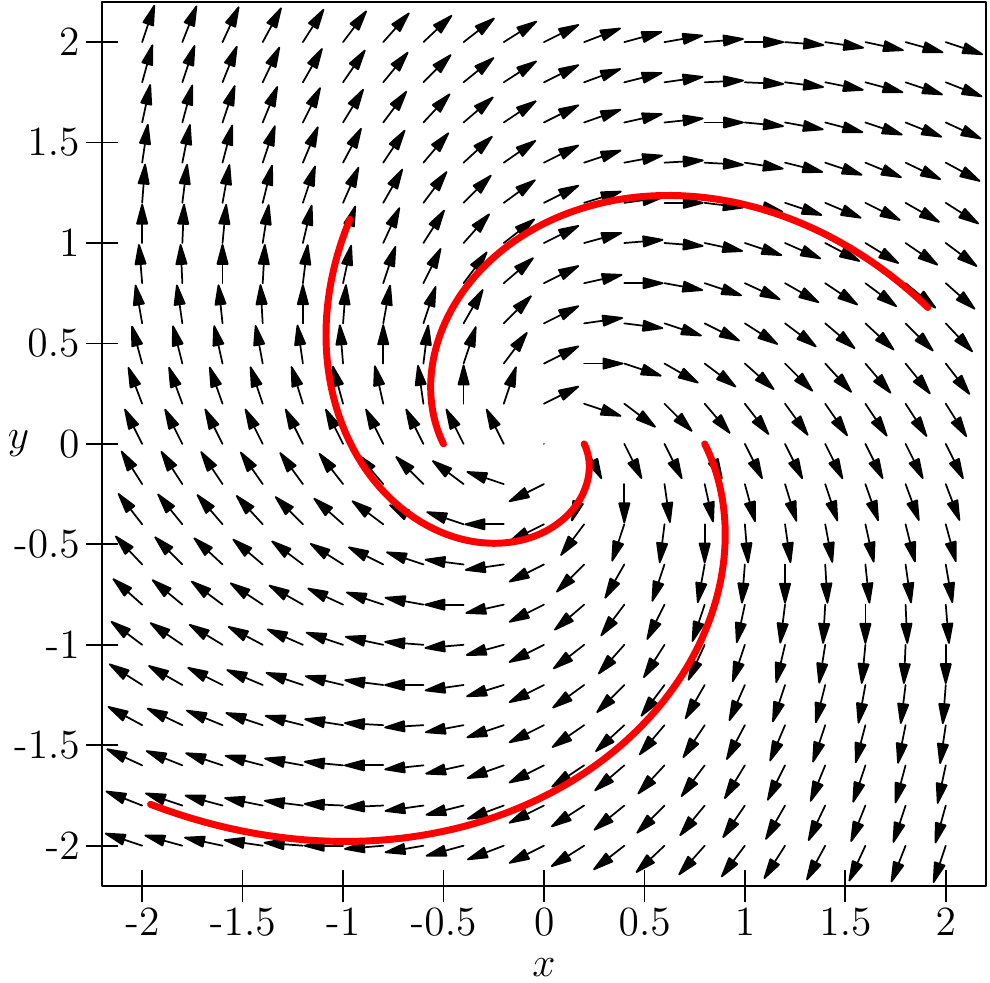}}\hfill
  \subfigure[center]{\includegraphics[width=0.325\textwidth]{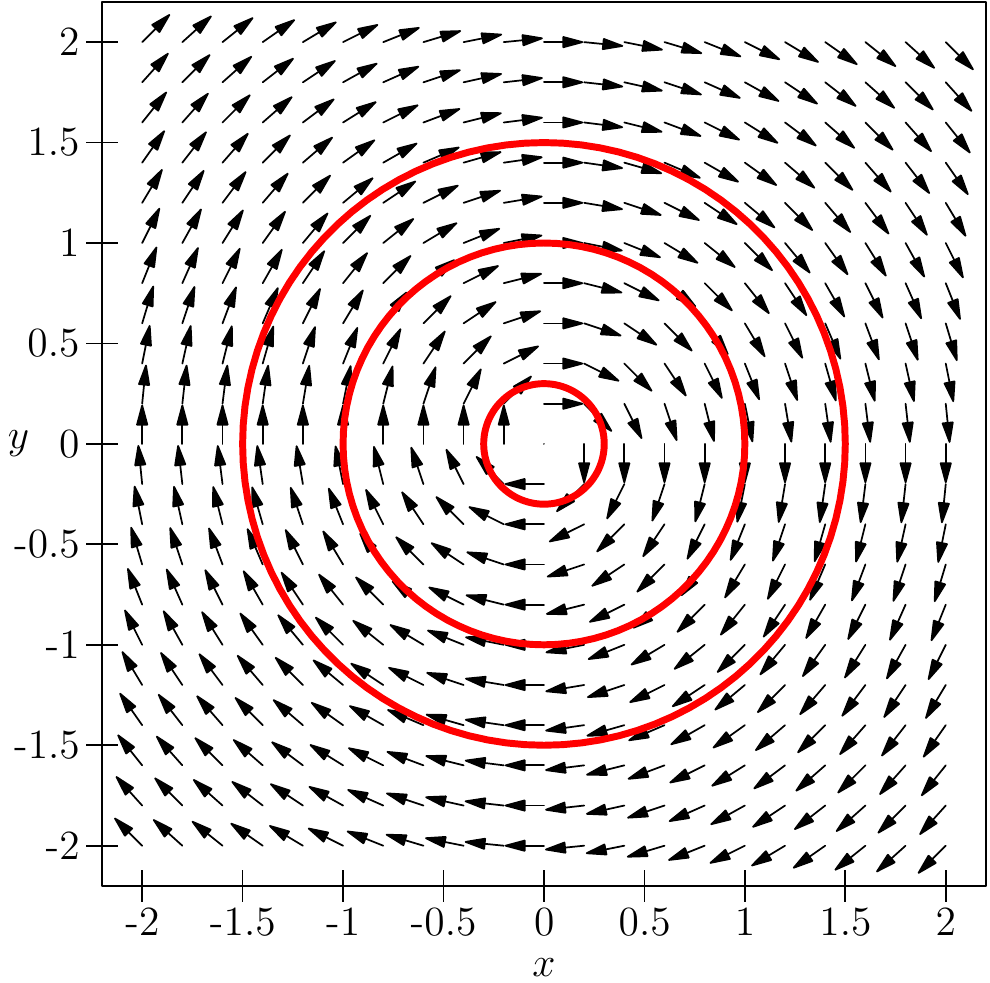}}\hfill
  \subfigure[stable focus]{\includegraphics[width=0.325\textwidth]{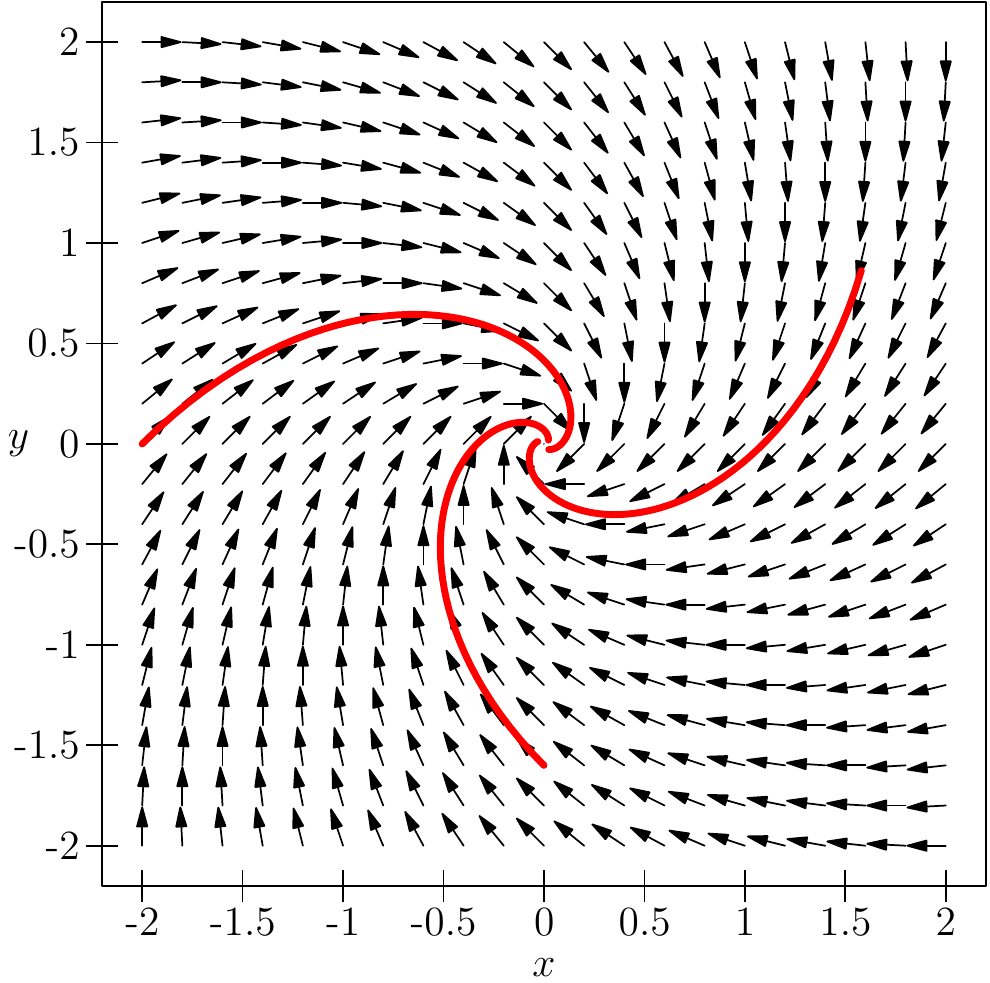}}
  \caption{Vector plots for critical points at $(x_c=0,y_c=0)$.
  The arrows are normalized to unit length.
  The red lines show some example orbits.}
  \label{fig:critPoints}
\end{figure}

Figure~\ref{fig:critPoints} shows examples of the six main types of critical points which are located at the centers of the vector field diagrams. The arrows are normalized to unit length to better see the characteristics of the field. The red curves represent some typical orbits.

To further characterize a vector field, one has to determine the characteristic lines that mark the boundary between different regions. These lines are also called separatrices, in general. A characteristic line is called \emph{homoclinic orbit} if the limits
\begin{equation}
  \lim\limits_{t\rightarrow\infty}\vec{\sigma}(t,\vec{x}_0) = \lim\limits_{t\rightarrow -\infty}\vec{\sigma}(t,\vec{x}_0) = \vec{x}_c
\end{equation}
have the same critical point $\vec{x}_c$. If the limits have different critical points, the characteristic line is called \emph{heteroclinic orbit}; otherwise, it is a generic \emph{orbit}.

\section{Visualization techniques}\label{sec:vistech}

\subsection{Vector field diagram}\label{subsec:vfd}
\begin{figure}[ht]
  \subfigure[]{\label{fig:vforig}\includegraphics[width=0.325\textwidth]{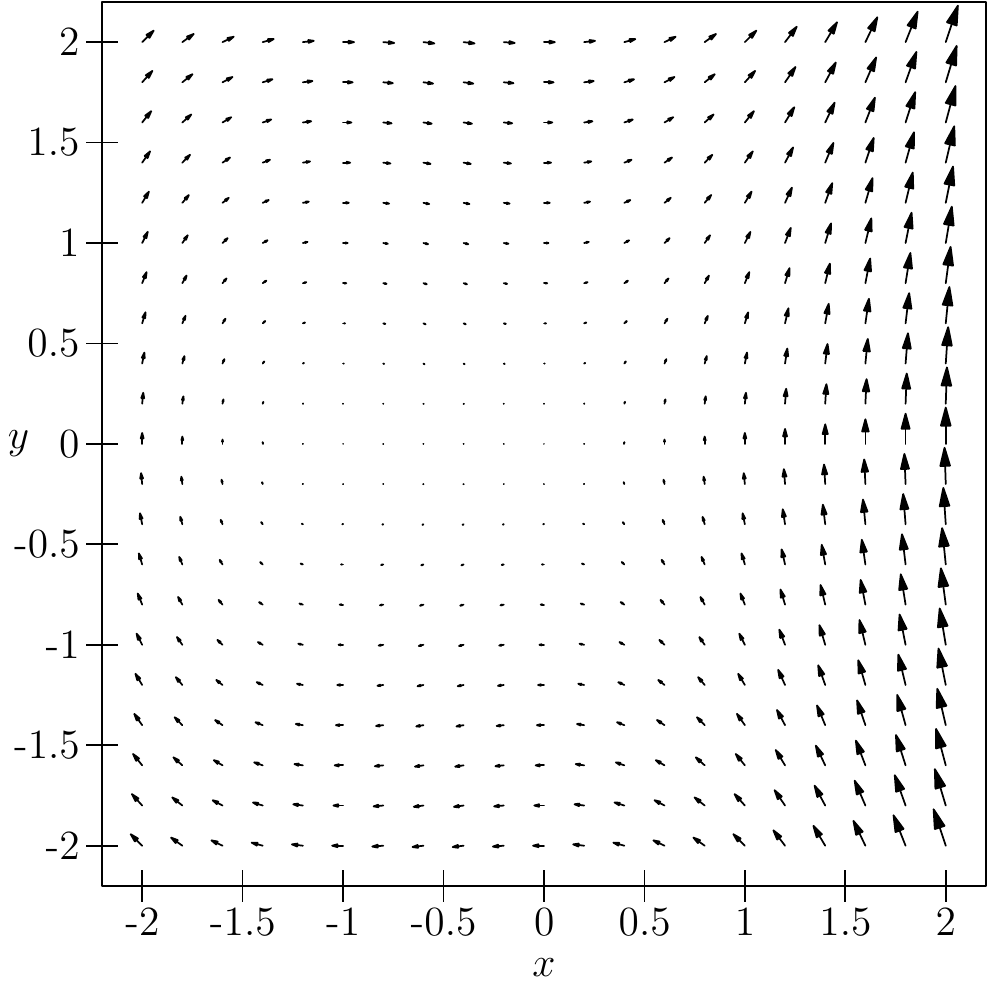}}\hfill
  \subfigure[]{\label{fig:vfnormalized}\includegraphics[width=0.325\textwidth]{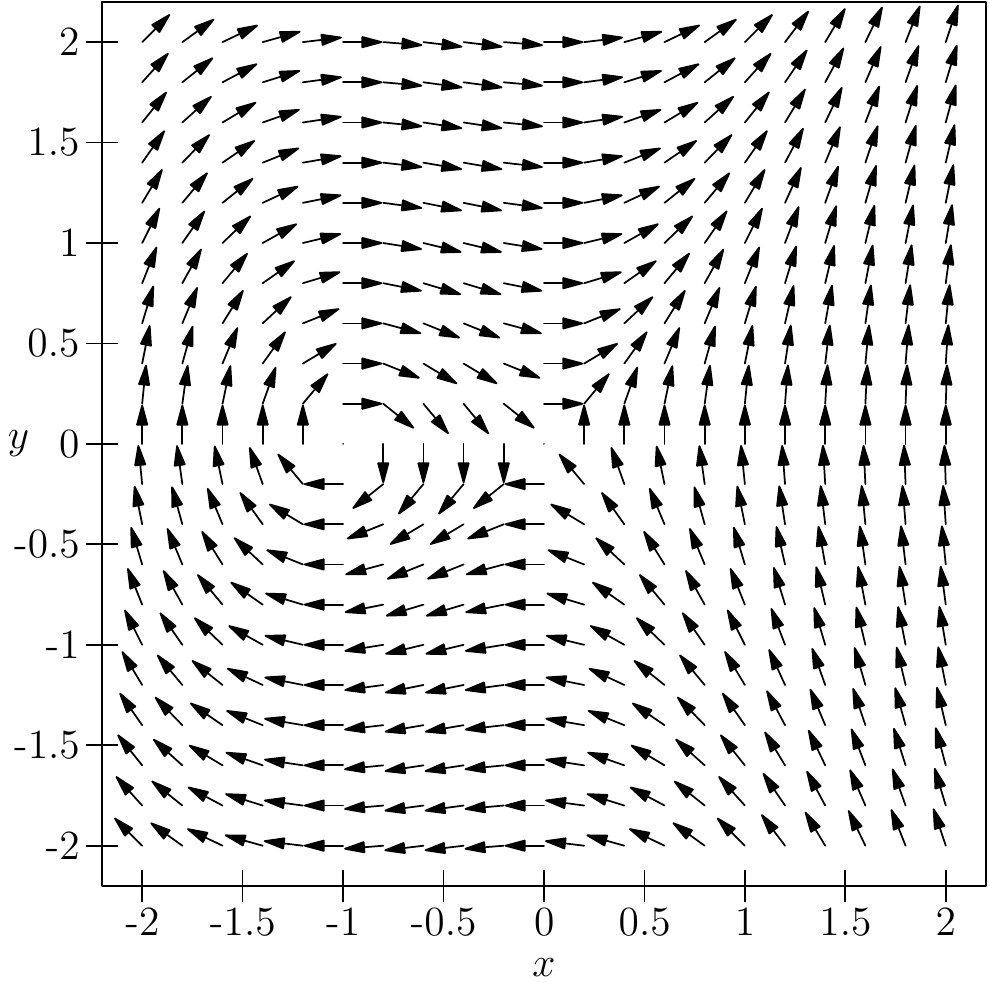}}\hfill
  \subfigure[]{\label{fig:vf2}\includegraphics[width=0.325\textwidth]{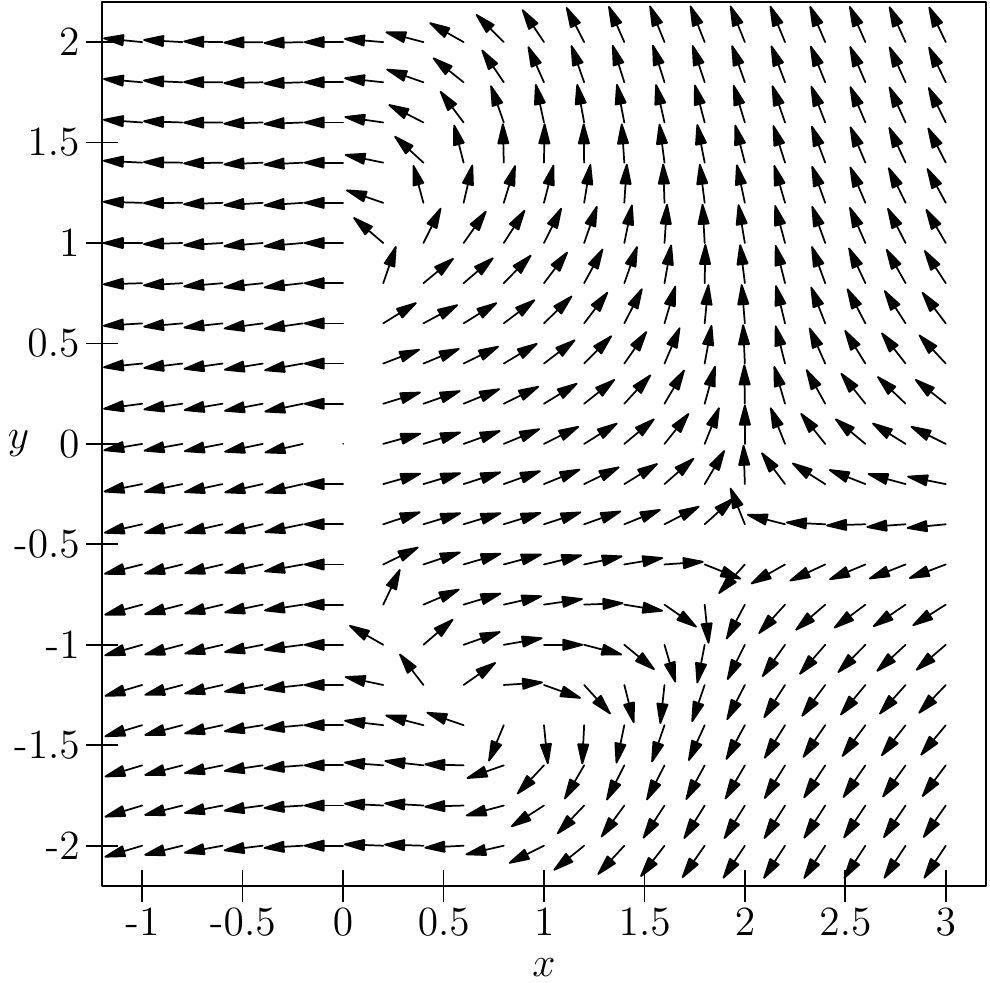}}
  \caption{Vector field diagram for $P(x,y)=y$, $Q(x,y)=x+x^2$ with critical points $\vec{c}_1=(-1,0)^T$ and $\vec{c}_2=(0,0)^T$; (a) real length, (b) normalized. 
  (c) Normalized vector field diagram for $P(x,y)=4x-2x^2-y^2$, $Q(x,y)=x(1+xy)$ with critical points $\vec{c}_3=(0,0)^T$, $\vec{c}_4=(0.734,-1.363)^T$, and $\vec{c}_5=(1.931,-0.518)^T$.}
  \label{fig:vecfield}
\end{figure}

The most straightforward visualization of the two-dimensional system (\ref{eq:system}) is a vector field diagram with arrows pointing in the direction $\vec{f}=(P,Q)^T$ as already shown in figure~\ref{fig:critPoints}.  
Figures~\ref{fig:vforig},\subref{fig:vfnormalized} show a vector field diagram where the arrows are either normalized or not.
Without normalization (Fig.~\ref{fig:vforig}), the arrow lengths are too diverse so that the overall structure of the vector field gets lost if the diagram cannot be scaled interactively.

While in figure~\ref{fig:vfnormalized} the critical points at $\vec{c}_1=(-1,0)^T$ and $\vec{c}_2=(0,0)^T$ are obviously a center and a saddle, respectively, the location and the types of the critical points in figure~\ref{fig:vf2}, $\vec{c}_3=(0,0)^T$, $\vec{c}_4=(0.734,-1.363)^T$, $\vec{c}_5=(1.931,-0.518)^T$, are not as obvious. 
Even more complicate is the retrieval of the separatrices in figure~\ref{fig:vf2}.

\subsection{Line integral convolution}\label{subsec:lic}
Line integral convolution is a well-known texture synthesis technique in flow visualization. 
It has the advantage that its spatial resolution is much higher than that of a vector field diagram. 
The idea is to locally smear an image  (usually an image consisting of white noise) along the direction of a vector field.
Thereby, pixel values along streamlines are strongly correlated while orthogonal to the vector field, there is virtually no correlation.

\begin{figure}[ht]
  \subfigure[]{\includegraphics[height=110px]{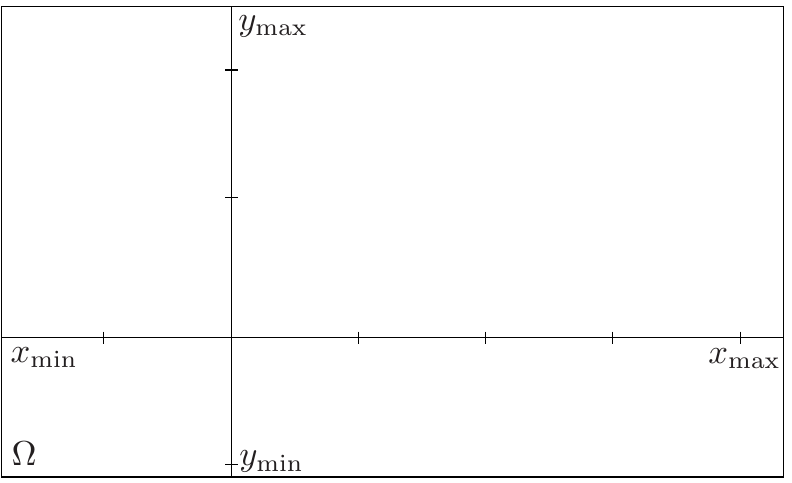}}\hfill
  \subfigure[]{\includegraphics[height=110px]{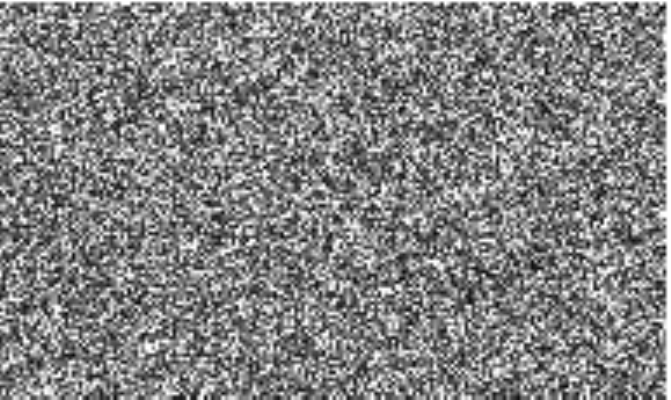}}
 \caption{(a) $2D$ domain of interest $\Omega=\{(x_{\mbox{\tiny min}},y_{\mbox{\tiny min}}),(x_{\mbox{\tiny max}},y_{\mbox{\tiny max}})\}$. (b) Noise texture/image $T$ with resolution $N_x\times N_y$ pixels, slightly smoothed by a Gaussian kernel to reduce high frequencies.}
 \label{fig:noiseTex}
\end{figure}

The basic algorithm works as follows. 
First, we have to generate a window-filling noise texture $T$ of size $N_x\times N_y$ pixels with random numbers in the range $[0,1]$ for each pixel that covers the domain $\Omega$, see figure~\ref{fig:noiseTex}. 

Next, we perform a discrete line integral convolution of the noise image for each pixel $\mathbf{p}\in T$ using the trajectory $\vec{\sigma}(t,\vec{x}_0)$.
The resulting intensity $I\left(\mathbf{p}\right)$ at the pixel $\mathbf{p}$ follows from
\begin{equation}
  I\left(\mathbf{p}\right) = k(0)T(\mathbf{p}) + \sum\limits_{n=1}^{N}\left[k(n)T(\mathbf{u}_n)+k(-n)T(\mathbf{v}_n)\right],
\end{equation}
where $k$ is a filter kernel and $2N+1$ is the filter length. 
In detail, we start with the intensity $T\left(\mathbf{p}\right)\in[0,1]$ of the noise texture at pixel $\mathbf{p}$ weighted by $k(0)$ and then sum up all the intensity values of $T$ along the positive and negative parts of the trajectory weighted by the kernel value $k$ at the corresponding sampling points $\mathbf{u}_n$ and $\mathbf{v}_n$, respectively."
For that, we first need the relation between pixel coordinates $\mathbf{q}=(q_x,q_y)\in T$ and domain coordinates $\vec{x}=(x,y)^T\in\Omega$ which is defined by
\begin{equation}
 \vec{x}\mapsto \mathbf{q}\quad\mbox{with}\quad q_x = N_x\frac{x-x_{\mbox{\tiny min}}}{x_{\mbox{\tiny max}}-x_{\mbox{\tiny min}}},\quad q_y = N_y\frac{y-y_{\mbox{\tiny min}}}{y_{\mbox{\tiny max}}-y_{\mbox{\tiny min}}}.
 \label{eq:coordRel}
\end{equation}
For the sake of convenience, we employ the Euler method to integrate the trajectories $\vec{\sigma}(t,\vec{u}_0=\vec{v}_0)$ in positive and negative direction. 
The initial pixel $\mathbf{p}$ defines the initial positions $\vec{u}_0=\vec{v}_0$ which follow from the inverse of equation (\ref{eq:coordRel}).
Then, Euler iteration delivers the sampling points
$\vec{u}_n,\vec{v}_n\in\Omega$ with
\begin{equation}
 \vec{u}_{n+1} = \vec{u}_n + h\vec{f}_n,\quad \vec{v}_{n+1} = \vec{v}_n - h\vec{f}_n,\qquad n=0,\ldots,N-1.
\end{equation}
According to equation (\ref{eq:coordRel}), we obtain the sampling points $\mathbf{u}_n,\mathbf{v}_n\in T$ from $\vec{u}_n$ and $\vec{v}_n$. 

The filter kernel for standard LIC can be any symmetric function. Here, we use a Hann filter
\begin{equation}
  \label{eq:hannFilter}
  k(n)=\frac{1}{2N}\left(1+\cos\frac{\pi n}{N}\right),\quad n=\{-N,\ldots,N\},\quad \sum\limits_{n=-N}^Nk(n)=1.
\end{equation}
The line integral convolution is computationally quite expensive. But with the high parallelism of GPUs, we can accomplish the convolution at interactive rates.
At the end, we obtain the LIC image composed of all the intensity values $I\left(\mathbf{p}\right)$ calculated above, which can afterwards be scaled arbitrarily to enhance the field structures, see, e.g., figure~\ref{fig:screenshot}.

\begin{figure}[ht]
  \includegraphics[height=175px]{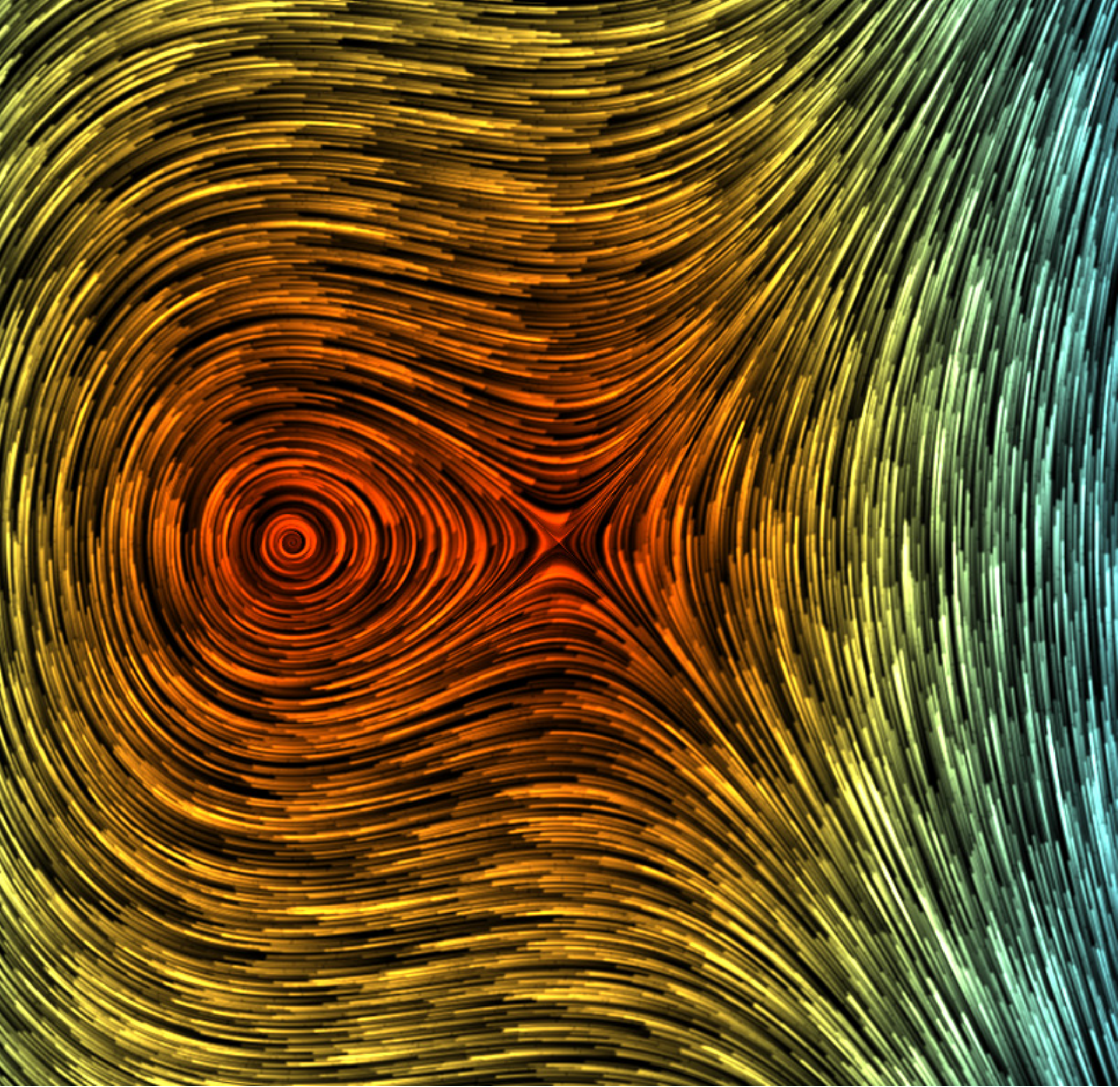}\hfill
  \includegraphics[height=175px]{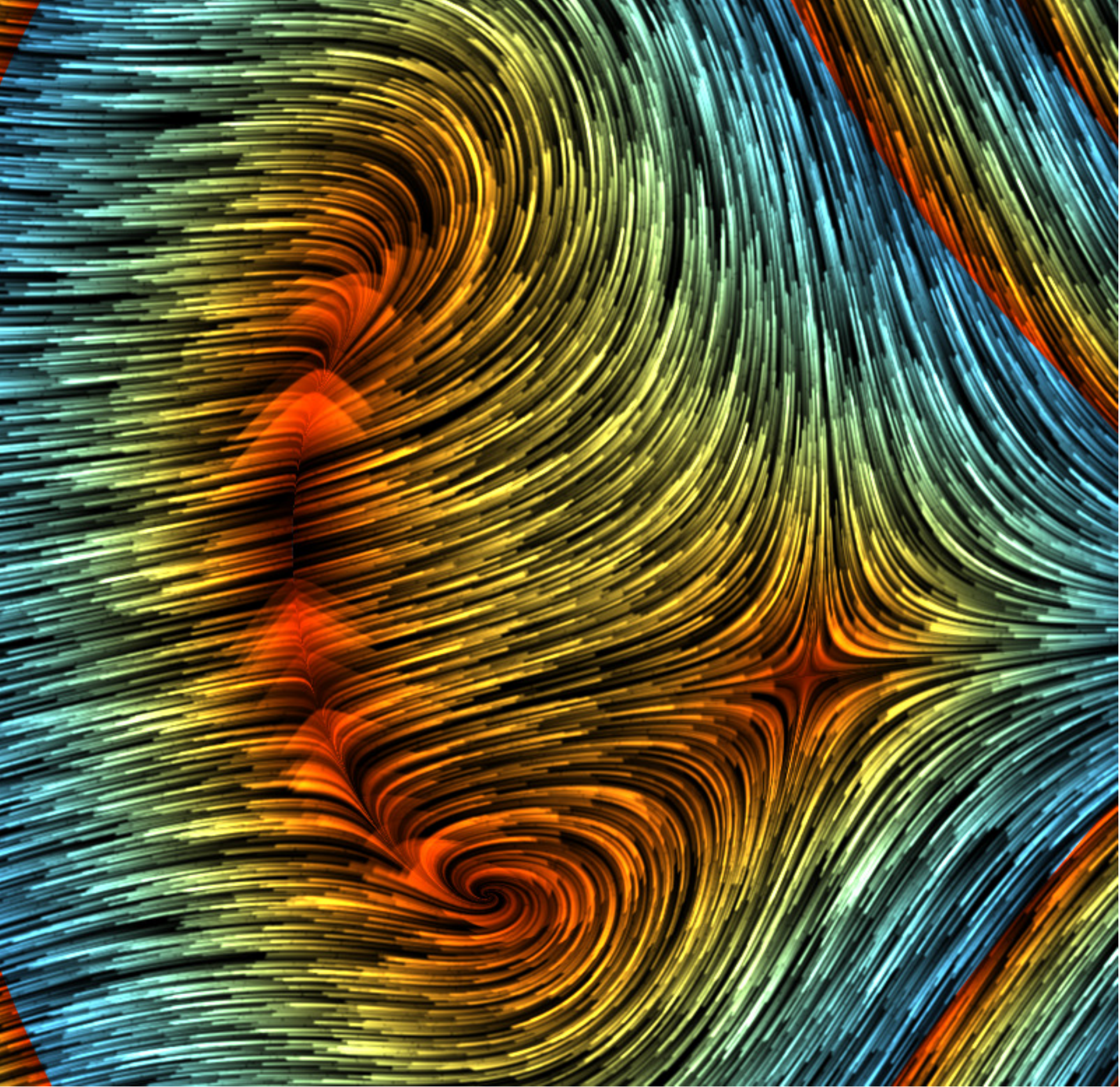}\\[0.1em]
  \includegraphics[width=1\textwidth]{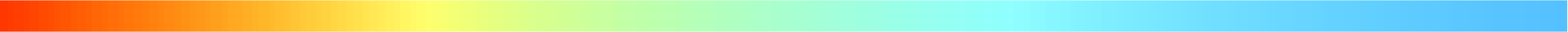}
  \caption{Oriented line integral convolution images of the vector fields of figure~\ref{fig:vecfield}, generated by \emph{ASysViewer} using the scripts {\tt paperExp1.js} and {\tt paperExp2.js}. Here, colour encodes vector magnitude $\|\vec{f}\|$. As these values are very diverse, we repeat the colour map with $n\in\mathbbm{N}$.}
  \label{fig:olic}
\end{figure}

Unfortunately, the standard LIC loses the directional information of the vector field, i.e., forward and reverse direction are indiscernible. 
To resolve this disadvantage, we can use an asymmetric filter kernel that comprises orientation (\emph{oriented LIC}), see, e.g., Wegenkittl et al.~\cite{Wegenkittl:1996}. 
The most straightforward kernel for that is a sawtooth. 
Furthermore, for oriented LIC, we have to replace the noise texture of figure~\ref{fig:noiseTex} by a texture with a finite number of randomly distributed spots and black background. 
Figure~\ref{fig:olic} shows the vector fields of figure~\ref{fig:vecfield} using oriented LIC. 
Additionally, we use a repeating colour map for the vector magnitudes $\|\vec{f}\|$.

Another possibility to maintain the direction information of the vector field is to animate the line integral convolution (\emph{animated LIC}). 
For that, we change the frequency of the Hann filter and modulate the phase by animation time $\tau$. 
Additionally, we multiply the modified Hann filter with a Hann window. 
Thus, $k(n)$ from equation (\ref{eq:hannFilter}) is replaced by
\begin{equation}
 h(n) = \frac{1}{4}\left[1+\cos\left(\frac{2\pi n}{N}+\tau\right)\right]\left(1+0.2\cos\frac{\pi n}{N}\right).
\end{equation}

\subsection{Topology extraction}\label{subsec:topoExtract}
So far, we have demonstrated the benefits of dense vector field visualizations by the LIC family of algorithms as compared to the discrete sparse visualizations by vector plots: due to their dense nature, they provide detailed information about the direction of $\vec{f}$, and to some extent also about its magnitude and orientation. 
However, it is still difficult to spot the critical points and, in particular, the separatrices emanating therefrom (figure~\ref{fig:olic}). 
In other words, it is still difficult to identify regions of qualitatively different behaviour.

Topology extraction usually starts with the extraction of critical points. 
A major difficulty with this step is, however, the lack of a theory on the existence and location of critical points in arbitrary (higher-order) vector fields. 
Nevertheless, in the special class of low-order vector fields arising from 2D (bilinear) tensor-product linear interpolation, their number is restricted to two. 
Hence, in vector fields where $\Omega$ is discretized into cells and $\vec{f}$ is discretized at the nodes (the corners of the cells), such a field is present in each cell due to bilinear interpolation. 

The algorithm to detect the critical points works as follows:
\begin{enumerate}
\item
  Detect cells with different signs at their nodes both in terms of $P$ and $Q$.
\item
  Such cells are candidates for stationary points. 
  However, a sationary point is present only if the zero-level isocontours of $P$ and $Q$ intersect.
\item
  Analytical derivation of these intersections is possible, but subject to numerical instability. 
  Thus, we follow a subdivision approach:
\item
  Each cell with different signs in $P$ and $Q$ is subdivided, and the resulting four children cells are again tested for different signs in $P$ and $Q$. 
  This procedure is repeated until no cell exhibits different signs in $P$ and $Q$ or until a maximum subdivision level (a prescribed accuracy) is achieved. 
  The centers of the remaining cells with different signs represent stationary points.
\item
  Those stationary points that have $\det(J)\neq 0$ are the desired critical points.
\end{enumerate}

Those critical points where $\det(J)< 0$ are of type saddle and hence give rise to separatrices. To escape the zero-velocity region of the critical point, for each of the four separatrices an offset step along the major (or minor) eigenvector of the Jacobian is applied and streamline integration in forward (or reverse) direction is carried out, resulting in the desired separatrices.


\section{The viewer}\label{sec:application}
\subsection{Graphical user interface}\label{subsec:gui}
\begin{figure}[ht]
   \centering
   \includegraphics[width=0.9\textwidth]{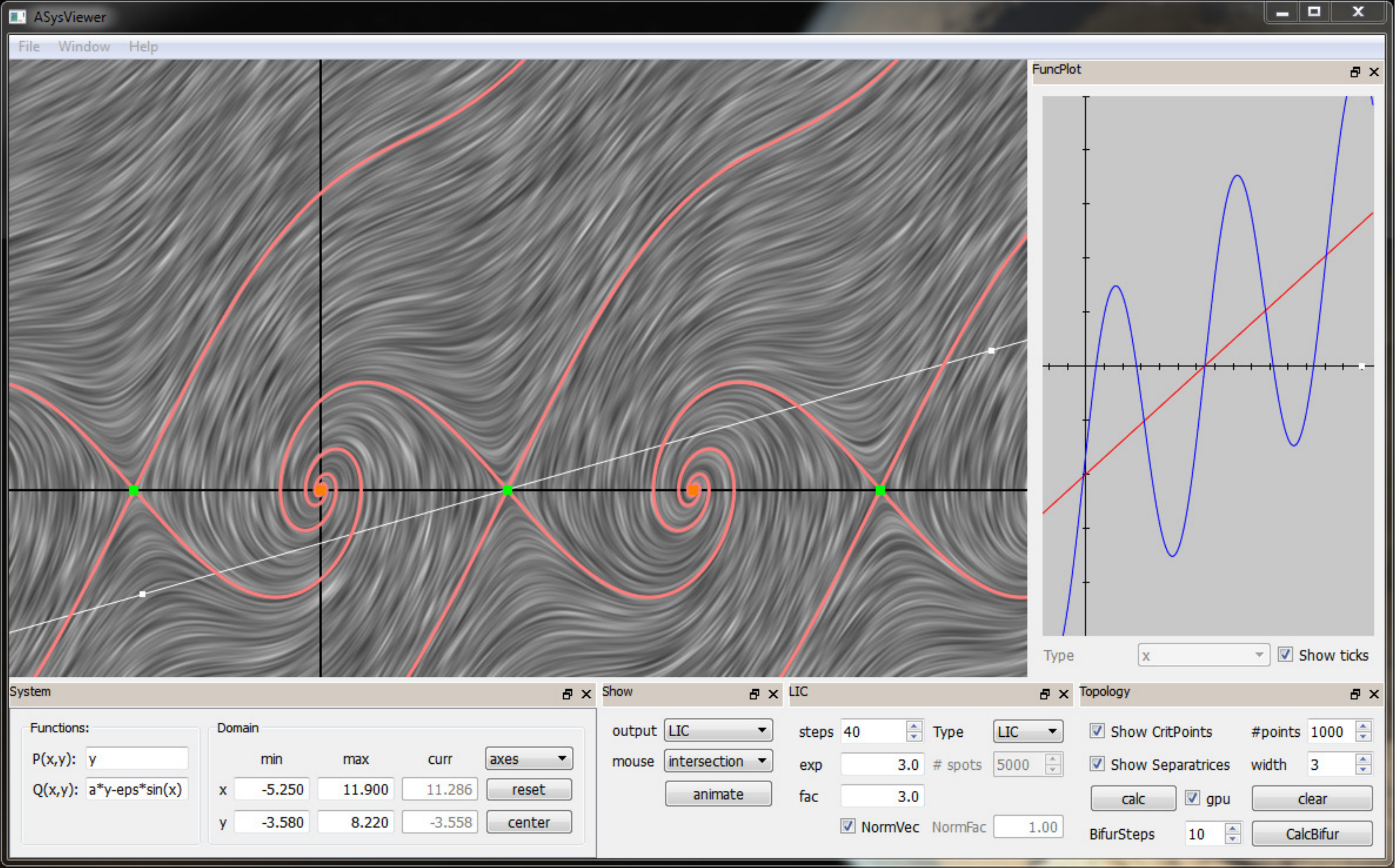}
   \caption{Screenshot of \emph{ASysViewer}'s graphical user interface showing standard LIC with separatrices and critical points. 
   In the ``{\tt FuncPlot}'' window, the blue and red lines correspond to the functions $P$ and $Q$ evaluated along the interactively defined white line in the main window.}
 \label{fig:screenshot}
\end{figure}

A screenshot of \emph{ASysViewer}'s graphical user interface is given in figure~\ref{fig:screenshot}. 
The main window shows standard LIC with critical points and separatrices of the ``mathematical pendulum with friction'' example, see Eq.~(\ref{eq:pendulum}). 
In the ``{\tt System}'' window, the functions $P(x,y)$ and $Q(x,y)$ can be given as mathematical expressions, and the domain $\Omega$ can be controlled either using mouse interaction (zooming: middle button; panning: left button) or by typing the values in the corresponding boxes. 
With the right mouse button a trajectory can be shown which starts at the current mouse position.
The parameters for this trajectory can be set within the ``{\tt Trajectory}'' window.
The ``{\tt FuncPlot}'' window shows the values of $P$ and $Q$ along the white intersection line drawn in the main window. 
This line can be set with the right mouse button and ``{\tt intersection}'' selected in the ``{\tt Show}'' window.
After defining the dynamical system, the critical points as well as the separatrices are determined by pressing the ``{\tt calc}'' button in the ``{\tt Topology}'' window.

\noindent There are also two windows to control the parameters {\tt a} and {\tt eps} which can be used as free parameters to define the functions $P$ and $Q$.

\subsection{Technical details}\label{subsec:tech}
\emph{ASysViewer} is based on the cross-platform application and user interface framework Qt~\cite{qt}, the graphics library OpenGL, and the OpenGL Shading Language (GLSL)~\cite{opengl}.
In the main window, we draw a single quad that covers the whole $2D$ domain of interest, see also figure~\ref{fig:noiseTex}.
For each fragment (pixel) of this quad, the line integral convolution is calculated within a fragment program/shader that works as a SIMD (single instruction multiple data) architecture.
The fragment program is written in a C-like programming language (GLSL) and is compiled and transferred to the GPU at runtime. 
When the functions $P$ and $Q$ are modified, the function strings are inserted in the fragment code and the fragment program is recompiled. 
Due to limitations of mathematical functions/operators in GLSL, expressions like ``$x^4$'' must be formulated as ``$x*x*x*x$''.

The integration of the trajectories to visualize the separatrices or individual orbits selected via mouse control is done using a step-controlled Runge-Kutta-Fehlberg integrator. 
For that, the functions $P$ and $Q$ are parsed using the \emph{muparser}~\cite{muparser} library by Ingo Berg.
For better reproducibility, the parameter settings of \emph{ASysViewer} are also scriptable by means of QtScript which is based on the ECMAScript~\cite{ecma} standard. 
The necessary scripting functions can be checked up from the source code.

\section{Examples}\label{sec:examples}
\subsection{Mathematical pendulum with friction}
As a first example, we consider the damped oscillation of the mathematical pendulum of unit length  which is described by the equation of motion $\ddot{\varphi} + \omega^2\sin\varphi + \alpha\dot{\varphi} = 0$ with deflection angle $\varphi$, angular frequency $\omega$, and a positive frictional coefficient $\alpha\geq 0$, see figure~\ref{fig:screenshot}. Substituting $\varphi=x$, $\dot{\varphi}=y$ converts the second order ordinary differential equation into the standard form of equation~\eref{eq:system},
\begin{equation}
  \dot{x} = P(x,y) = y,\qquad \dot{y} =  Q(x,y) = -\omega^2\sin x - \alpha y.
  \label{eq:pendulum}
\end{equation}
The positions of the corresponding critical points immediately follow from $\dot{x}=0=\dot{y}$. Thus, $y_c=0$ and $x_c=\sin(n\pi)$ with $n\in\mathbbm{Z}$. The Jacobian yields $\rm{tr}(J)=\alpha$ and $\det(J)=\omega^2\cos(n\pi)$. If $n$ is odd, the determinant is negative and the critical point is a saddle. If $n$ is even, on the other hand, the critical point is either a center (undamped oscillator, $\alpha=0$), an unstable focus ($\alpha^2<4\omega^2\cos(n\pi)$), or an unstable node. 
In the damped case, each unstable focus/node is connected to the two neighbouring saddle points via heteroclinic orbits, whereas in the undamped case, each saddle point is connected to its two neighbouring saddle points.

\subsection{Limit cycle}
A \emph{limit cylce} is an isolated periodic solution which is defined by the set of all points $\vec{y}$ with $\vec{\sigma}(t,\vec{x})\rightarrow\vec{y}$ for $t\rightarrow\infty$. An example can be found in the phase space for the differential equation related to the oscillation of a violin string as derived by Rayleigh~\cite{rayleigh1877},
\begin{equation}
    \ddot{v} - \mu\left(1-\frac{1}{3}\dot{v}^2\right)\dot{v}+v=0,\qquad \mu>0,
    \label{eq:rayleigh}
\end{equation}
which can be brought into the standard form (Eq.~(\ref{eq:system})) via $v=x$ and $\dot{v}=y$.
\begin{figure}[ht]
  \subfigure[]{\label{fig:rayleighOLIC}\includegraphics[width=0.48\textwidth]{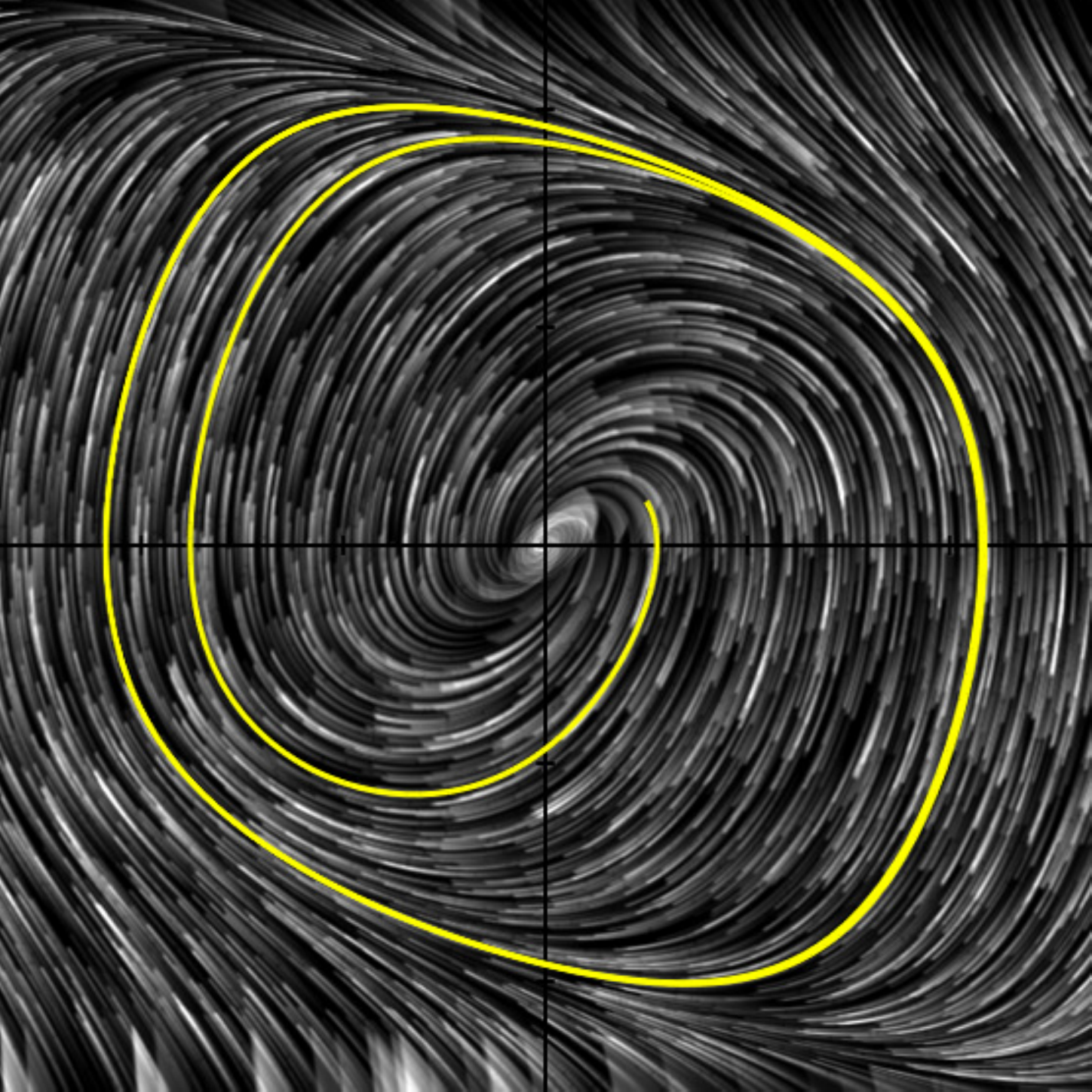}}\hfill
  \subfigure[]{\label{fig:rayleightVF}\includegraphics[width=0.48\textwidth]{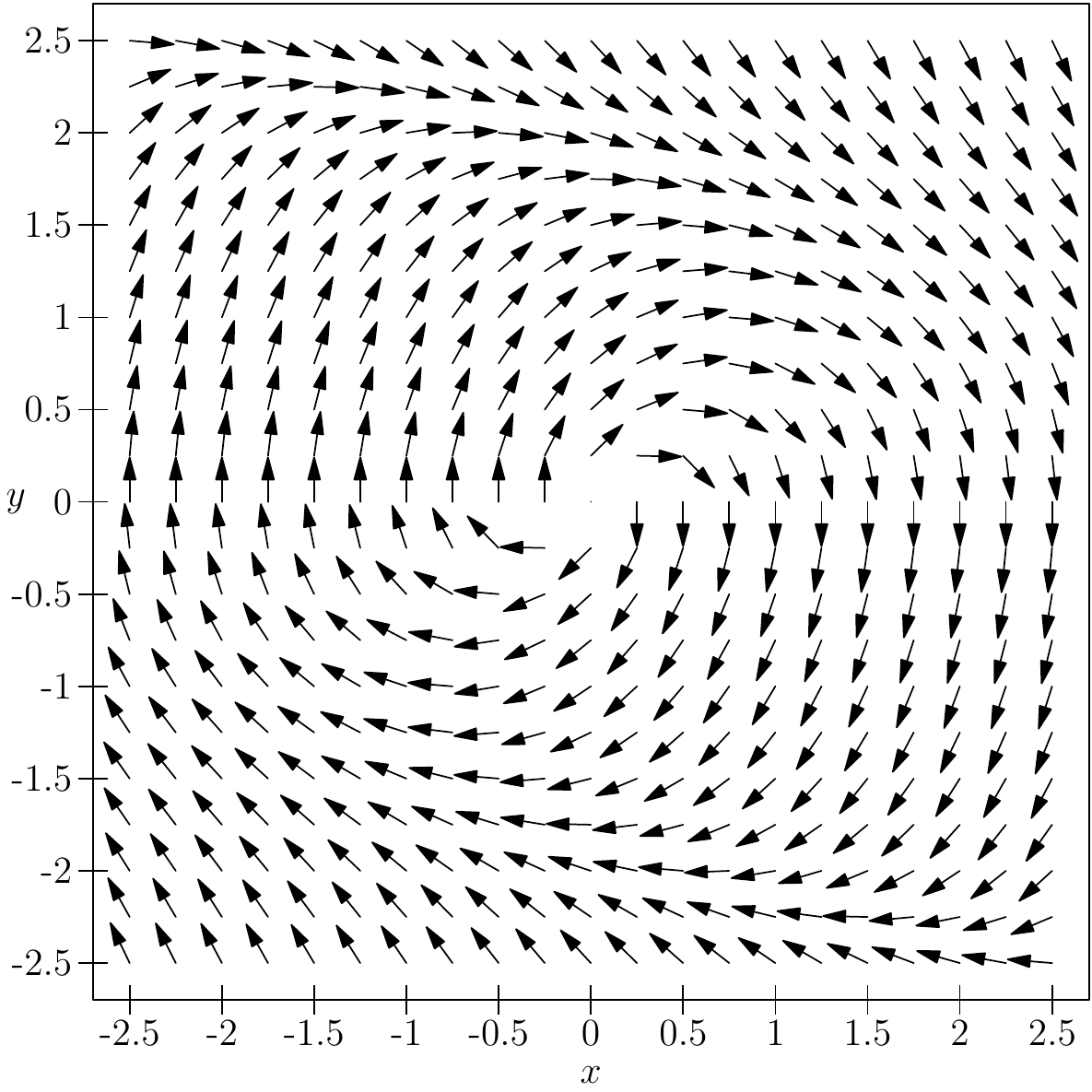}}
  \caption{(a) Phase space image of the Rayleigh system, Eq.~(\ref{eq:rayleigh}), visualized using oriented LIC. The trajectory (yellow line) starting at $x=0.5, y=0.2$ approaches the limit cycle asymptotically. Here, $\Omega=[-2.5,2.5]\times[-2.5,2.5]$ and $\mu=1$. (b) Vector field diagram.}
  \label{fig:rayleigh}
\end{figure}

As shown in the phase space image, figure~\ref{fig:rayleighOLIC}, the oriented LIC representation of the Rayleigh system indicates that there must exist a boundary curve separating the outflowing and inflowing vector field. Every integral curve $\vec{\sigma}(t,\vec{x}_i)$ approaches this boundary curve (limit cycle) irrespective of the initial point $\vec{x}_i$.

\subsection{Poincar{\'e} index}
The \emph{Poincar{\'e} index} $i_P$ of a closed curve $\Gamma$ is determined by integrating the change in the angle of the vectors at each point of $\Gamma$ along $\Gamma$. 
For the numerical calculation, we set $\Gamma$ to a circle of radius $r$ and replace the integration by the sum
\begin{equation}
 i_p = \frac{1}{2\pi}\sum\limits_{n=0}^{N}\arcsin\frac{\|\vec{f}_n\times\vec{f}_{n+1}\|}{\|\vec{f}_n\|\cdot\|\vec{f}_{n+1}\|},
\end{equation}
where $\vec{f}_n=\vec{f}(r\cos\varphi_n,r\sin\varphi_n)^T$ and $\varphi_n=2\pi n/N$. 
For a sink, source, or center, we have $i_P=+1$. 
A saddle point has $i_P=-1$. 
For details to the Poincar{\'e} index, we refer the reader to Wiggins~\cite{wiggins}.
In \emph{ASysViewer}, the Poincar{\'e} index of an arbitrary point $\vec{x}$ can be determined interactively by constructing a circle around $\vec{x}$ via mouse handling (right button). For that, select ``{\tt poincare idx}'' in the mouse combo box of the ``{\tt Show}'' window.

\subsection{Pitchfork bifurcation}
Consider the system
\begin{equation}
  \dot{x} = ax-x^3,\qquad \dot{y} = -y,
  \label{eq:pitchfork}
\end{equation}
with free parameter $a\in\mathbb{R}$. 
If $a$ is continuously varied from negative to positive values, this system undergoes a \emph{pitchfork bifurcation}.
As long as $a$ is negative, there is only a single critical point at $\vec{x}_{neg}=(0,0)^T$. For $a=0$, there is one non-hyperbolic critical point at the origin. 
If $a>0$, we obtain three critical points $\vec{x}_1=(0,0)^T$ (saddle), $\vec{x}_2=(-\sqrt{a},0)^T$ (stable node), and $\vec{x}_3=(\sqrt{a},0)^T$ (stable node).

\begin{figure}[ht]
  \subfigure[``pos,neg $P,Q$'' image for $a=1$]{\includegraphics[height=175px]{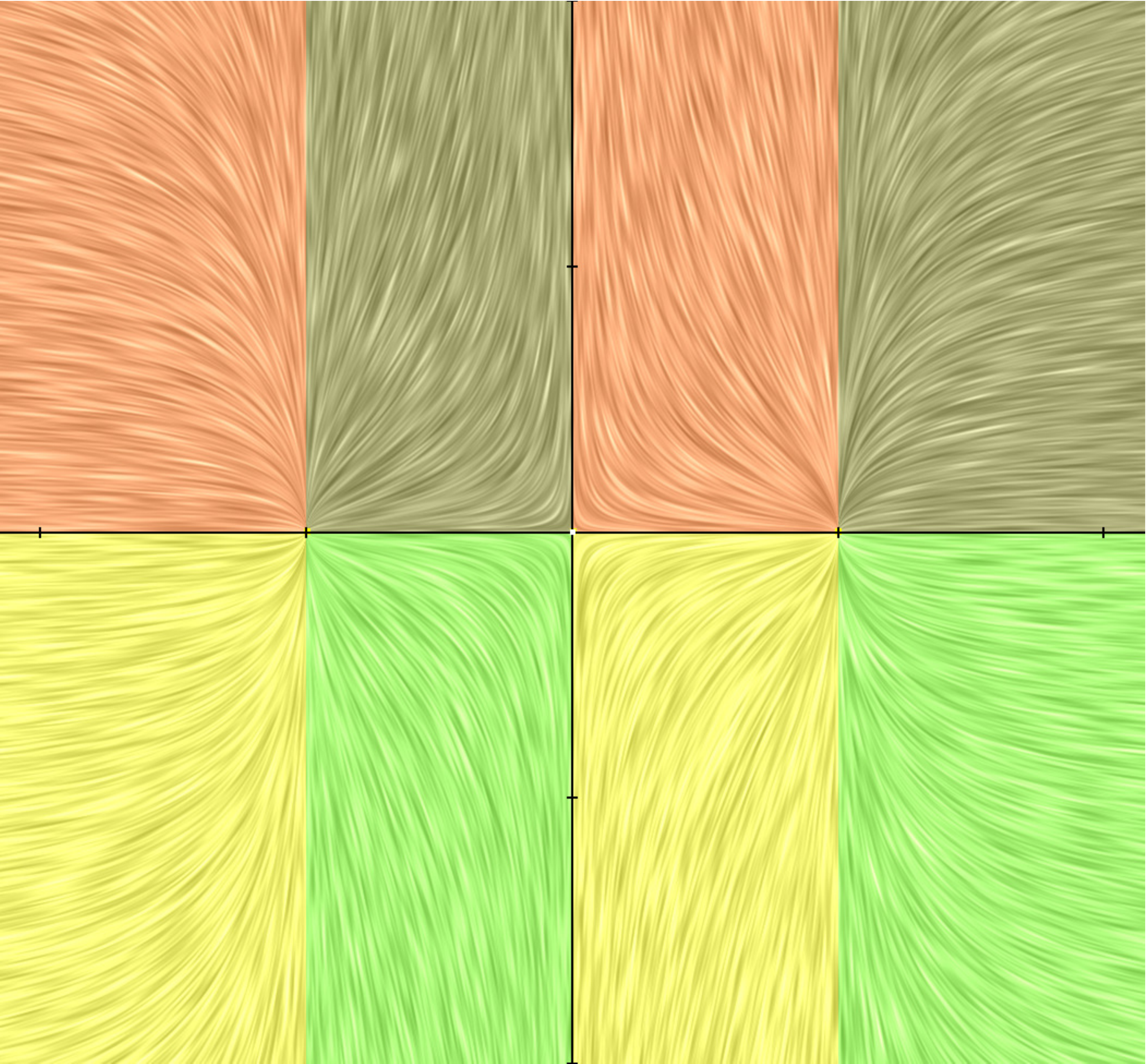}}\hfill
  \subfigure[Bifurcation diagram]{\includegraphics[height=175px]{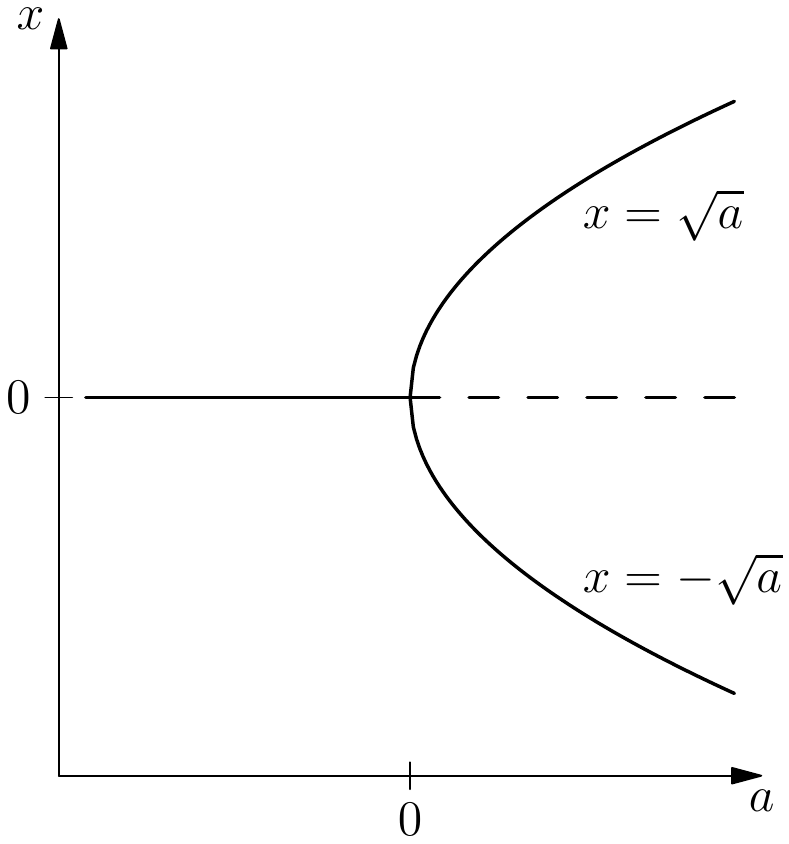}}
  \caption{The system (Eq.~(\ref{eq:pitchfork})) shows a pitchfork bifurcation at $a=0$.
  If $a>0$, the critical point at the origin is a saddle whereas the other two are stable nodes.
  The colour coding in (a) is as follows: yellow ($P>0,Q>0$), red ($P>0,Q<0$), brown ($P<0,Q<0$), green ($P<0,Q>0$); (b) pitchfork bifurcation.}
  \label{fig:pitchfork}
\end{figure}

Figure~\ref{fig:pitchfork} shows the ``pos,neg $P,Q$'' image for the pitchfork bifurcation system with $a=1$, where colour indicates the signs of the functions $P$ and $Q$.
Hence, a point where all four colours hit is a stationary point, $P=Q=0$.

%
%
%

%
%
%

\section{Summary}
In this work we presented a framework for interactive visual exploration of $2D$ autonomous dynamical systems.
For that, we use different types of line integral convolution techniques for a dense representation of the vector field.
We also described a robust algorithm to numerically detect critical points.
As the source code of \emph{ASysViewer} is freely available, the interested student is encouraged to conduct his or her own experiments and to look also at the coding details.


\ack
This work was partially funded by Deutsche Forschungsgemeinschaft (DFG) as part of the Collaborative Research Centre SFB 716 and the Cluster of Excellence in Simulation Technology (EXC 310) at the University of Stuttgart.

\section*{References}
\bibliographystyle{unsrt}
\bibliography{lit_asys}

\end{document}